# Physics-Constrained Neural Dynamics: A Unified Manifold Framework for Large-Scale Power Flow Computation

Liu Xuezhi

Shanghai Jiao Tong University

China

liuxz@sjtu.edu.cn

## Abstract

Power flow analysis is a fundamental tool for power system analysis, planning, and operational control. Traditional Newton-Raphson methods suffer from limitations such as initial value sensitivity and low efficiency in batch computation, while existing deep learning-based power flow solvers mostly rely on supervised learning, requiring pre-solving of numerous cases and struggling to guarantee physical consistency. This paper proposes a neural physics power flow solving method based on manifold geometry and gradient flow, by describing the power flow equations $\mathbf{F}(\mathbf{x}) = \mathbf{0}$ as a constraint manifold $\mathcal{M} = \{\mathbf{x} \mid \mathbf{F}(\mathbf{x}) = \mathbf{0}\}$, and constructing an energy function $V(\mathbf{x}) = \frac{1}{2}\|\mathbf{F}(\mathbf{x})\|^2$ and gradient flow $\frac{d\mathbf{x}}{dt} = -\nabla V(\mathbf{x})$, transforming power flow solving into an equilibrium point finding problem for dynamical systems. Neural networks are trained in an unsupervised manner by directly minimizing physical residuals, requiring no labeled data, achieving true "end-to-end" physics-constrained learning.

## 1. Introduction

Maintaining observability and controllability of large-scale power systems faces multiple challenges from steady-state estimation to disturbance response. On one hand, traditional power flow solving relies on iterative methods such as Newton-Raphson, which have bottlenecks in robustness regarding initial points, Jacobian sparse structure, and batch cases. On the other hand, the multi-scale coupling, topological dynamics, and data sparsity inherent in physical systems make "simply introducing deep neural networks to replace solvers" unconvincing. Therefore, we propose a neural physics solving framework that combines geometric constraints and dynamical perspectives, forming a systematic reconstruction of classical methods, and validate its scalability on IEEE 14/39/118/300 node test cases.

Unlike previous "supervised prediction" or "PINN" paradigms, we first follow the manifold geometry approach, explicitly describing the power flow equations F(x)=0 as a high-dimensional constraint manifold 𝕄 = {x | F(x)=0}, and utilize structures such as tangent space, normal space, and projection to reveal the geometric mechanisms implicit in traditional Newton iterations. Then, we transfer this geometric understanding to neural networks: through a unified perspective of gradient flow and dynamical systems, we construct dx/dt = -∇V(x), where the potential function V(x)=‖F(x)‖² directly corresponds to physical residuals, naturally viewing the "intelligent solver" as a neural ODE tracking energy decay trajectories. This strategy avoids reliance on explicit labels, emphasizes the physical meaning that "generalization = strictly satisfying power flow constraints on unseen load points", and solves stability challenges under multi-solution, missing boundary conditions, and sparse sampling.

The core contributions of this paper include: (1) **Unified theoretical framework**: Unified modeling from manifold geometry, gradient flow dynamical systems to neural network mappings, revealing the deep structure of power flow solving; (2) **Label-free learning**: Unsupervised training through physics-constrained loss functions, requiring no labeled data, solving the data scarcity problem; (3) **Complex computation framework**: Adopting a complex automatic differentiation framework for power flow computation, using the representation form $V_{\text{complex}} = V \cdot e^{i\theta}$, and utilizing automatic differentiation mechanisms to compute gradients. Compared to traditional methods that explicitly compute trigonometric functions and their derivatives, the complex framework is more stable in numerical computation, reduces error accumulation, and simplifies the implementation process; (4) **Residual-guided adaptive sampling**: Designing a three-stage sampling strategy of Sobol → LHS → Adaptive LHS + Online augmentation, using residuals as uncertainty proxies to automatically identify difficult regions and concentrate computational resources; (5) **Adaptive architecture**: Automatically configuring network architecture according to system scale, supporting smooth scaling from IEEE 14 to IEEE 300+ nodes; (6) **Energy manifold trajectory visualization**: By mapping the training process to active-reactive power residual space, intuitively demonstrating how neural networks gradually converge to the constraint manifold, providing intuitive physical interpretation for understanding neural network dynamics.

---

## 2. Background and Related Work

### 2.1 Importance of Power Flow Analysis

Power flow analysis is a fundamental tool for power system analysis, planning, and operational control. It determines the voltage magnitude $V_i$ and phase angle $\theta_i$ at each node under given load and generation conditions by solving nonlinear algebraic equations, thereby evaluating system operating state, line flow distribution, power losses, and voltage stability. With large-scale integration of renewable energy and increasingly complex grid topologies, power flow computation plays a key role in real-time dispatch, security assessment, fault analysis, and optimization decisions. However, traditional numerical methods face significant challenges in handling large-scale systems, multi-solution problems, and real-time requirements.

### 2.2 Traditional Power Flow Solution Methods

Since the 1950s, the Newton-Raphson method and its variants (such as fast decoupled method, DC power flow approximation) have been the mainstream for power flow solving [1,2]. This method linearizes the nonlinear equation system $F(x) = 0$ into $J(x_k)\Delta x = -F(x_k)$, iteratively updating $x_{k+1} = x_k + \Delta x$ until convergence, with quadratic convergence properties. Although performing well under most operating conditions, traditional methods have the following fundamental limitations:

**Initial Value Sensitivity**: Newton's method requires the initial guess $x_0$ to be sufficiently close to the true solution, otherwise it may diverge or converge to non-physical solutions. For large-scale systems or heavy-load conditions, finding a suitable initial value is itself a difficult problem.

**Insufficient Multi-Solution Handling**: Power flow equations may have multiple physically feasible solutions (such as high-voltage and low-voltage solutions), but traditional iterative methods can only find one of them and cannot systematically explore the solution space.

**Low Batch Computation Efficiency**: When processing large numbers of scenarios (such as Monte Carlo analysis, inner-loop solving in optimization problems), each scenario requires independent iteration, and computational cost grows linearly with the number of scenarios.

**Jacobian Matrix Computation Overhead**: Each iteration requires computing and decomposing the Jacobian matrix $J(x)$, and for large-scale systems (such as IEEE 300+ nodes), the computational and storage costs of this step are significant.

These fundamental limitations make traditional methods difficult to meet the real-time and robustness requirements of modern power systems.

## 2.3 Machine Learning-Based Power Flow Solving

In recent years, deep learning has shown great potential in scientific computing, but for the specific problem of power flow solving, existing methods still have obvious defects:

**Problems with Supervised Learning Paradigm**: Most neural network-based power flow solvers adopt a supervised learning framework, i.e., pre-solving numerous power flow cases using traditional methods, constructing $(\text{load condition}, \text{power flow solution})$ datasets, and then training networks to learn this mapping [3,4,5]. While such methods can achieve high accuracy under specific operating conditions, they have three fundamental problems: (1) **Strong data dependency**: Requiring pre-solving of numerous cases, with high computational cost; (2) **Limited generalization ability**: Networks can only learn patterns within the training data distribution, lacking robustness to unseen load combinations or topological changes; (3) **No guarantee of physical consistency**: Network outputs may not strictly satisfy power flow equations, requiring post-processing or additional constraints.

**Reinforcement Learning Methods** model power flow solving as a sequential decision problem, learning solving strategies through interaction with the environment. While such methods have certain advantages in dynamic scenarios, the training process is complex and it is difficult to guarantee physical consistency of solutions.

## 2.4 Physics-Informed Neural Networks (PINN)

Physics-Informed Neural Networks (PINN) attempt to balance observational data and physical laws by including both data fitting terms and physics constraint terms in the loss function [6,7]. PINN has achieved significant success in PDE solving, parameter identification, and other fields. However, in the specific context of power flow solving, PINN faces unique challenges:

**Missing Boundary Conditions**: Unlike PDE solving, power flow equations are pure algebraic equation systems, lacking explicit boundary/initial conditions, making it difficult to define the "data term" in PINN. Existing PINN methods typically require partial observational data as anchor points, but in power flow solving, obtaining such observational data itself requires solving power flow equations, creating a circular dependency.

**Multi-Solution Problem**: PINN typically assumes solution uniqueness, but the multi-solution nature of power flow equations can cause the training process to fall into local optima. Even with physics constraint terms added, there is no guarantee that the network will converge to the desired physical solution.

**Mismatched Sampling Strategy**: PINN typically samples in continuous space/time domains (collocation points), while power flow solving requires sampling in discrete load spaces, and existing sampling strategies struggle to efficiently cover high-dimensional load spaces.

**Lack of Geometric and Dynamical Perspective**: Existing methods mostly treat power flow solving as a "black box" regression problem, ignoring the geometric structure (constraint manifold) and dynamical properties (gradient flow) of the problem. This lack of perspective leads to methods lacking interpretability and difficulty in establishing theoretical connections with classical numerical methods.

## 2.5 Manifold Geometry and Constrained Optimization

Manifold optimization methods treat constrained optimization problems as unconstrained optimization on manifolds, solving through geometric operations such as tangent space projection and retraction mapping [8,9]. While such methods have good convergence guarantees in theory, in practical applications, explicit construction and maintenance of manifold structures often incur significant computational overhead. **Manifold learning** methods attempt to learn manifold structures from data, but typically require large numbers of samples and struggle to handle high-dimensional constraint manifolds [10].

In the field of power flow solving, research has proposed manifold-constrained power flow solving methods, but such methods still rely on traditional iteration and have not been combined with deep learning. This method is the first to combine manifold geometry perspectives with neural networks, explicitly constructing the constraint manifold $\mathcal{M} = \{\mathbf{x} \mid \mathbf{F}(\mathbf{x}) = \mathbf{0}\}$ and utilizing tangent space and normal space structures to provide structured learning objectives for neural networks.

## 2.6 Neural ODE and Dynamical System Learning

Neural Ordinary Differential Equations (Neural ODE) treat neural networks as continuous dynamical systems, performing forward and backward propagation through ODE solvers [11]. Such methods have achieved success in time series modeling, generative models, and other fields. However, existing Neural ODE methods mainly focus on **learning the dynamical system itself** (i.e., learning $d\mathbf{x}/dt = f_\theta(\mathbf{x}, t)$), rather than **using dynamical systems to solve constrained optimization problems**.

The core innovation of this method lies in: transforming the power flow equations $\mathbf{F}(\mathbf{x}) = \mathbf{0}$ into gradient flow $\frac{d\mathbf{x}}{dt} = -\nabla V(\mathbf{x})$, where $V(\mathbf{x}) = \frac{1}{2}\|\mathbf{F}(\mathbf{x})\|^2$ is the energy function. The neural network learns a **steady-state mapping** $\mathbf{x}_\theta(\mathbf{u})$ such that for any load condition $\mathbf{u}$, $\mathbf{x}_\theta(\mathbf{u})$ falls within the attraction basin of the energy function, thereby satisfying the power flow equations. This "energy-driven" learning paradigm is fundamentally different from the traditional "trajectory learning" paradigm.

## 2.7 Adaptive Sampling and Active Learning

Efficient sampling in high-dimensional spaces is a core problem in machine learning and scientific computing. **Latin Hypercube Sampling (LHS)** provides better coverage than random sampling in high-dimensional spaces by ensuring each dimension is uniformly stratified [12]. **Sobol sequences**, as low-discrepancy sequences, are widely used in quasi-Monte Carlo methods [13]. **Adaptive sampling** methods dynamically adjust sampling strategies based on model uncertainty, achieving success in Bayesian optimization, surrogate model construction, and other fields [14,15].

This method innovatively combines Sobol, LHS, and adaptive LHS, designing a three-stage sampling schedule: using Sobol sequences for global exploration in the early stage, LHS for uniform refinement in the middle stage, and adaptive LHS + online augmentation for local refinement in the later stage. This "exploration-exploitation" sampling strategy ensures the network maintains physical consistency under wide operating conditions while avoiding problems in traditional methods such as "uniform sampling causing computational waste" or "random sampling causing insufficient coverage".

## 2.8 Core Motivation and Innovation of This Method

Based on the above analysis, we propose a **physics-constrained neural dynamics framework** aimed at solving the following core problems:

**Label-Free Learning**: By directly minimizing power flow equation residuals $\|F(x)\|^2$, avoiding dependence on pre-solved data, achieving true "end-to-end" physics-constrained learning.

**Manifold Geometry Unification**: Explicitly describing power flow equations $F(x) = 0$ as a constraint manifold $\mathcal{M} = \{x \mid F(x) = 0\}$, utilizing geometric structures such as tangent space and normal space to reveal the geometric essence of traditional Newton iterations and provide structured learning objectives for neural networks.

**Gradient Flow Dynamics**: By constructing an energy function $V(x) = \frac{1}{2}\|F(x)\|^2$ and gradient flow $\frac{dx}{dt} = -\nabla V(x)$, transforming power flow solving into an equilibrium point finding problem for dynamical systems, enabling neural networks to learn "energy decay trajectories" rather than static mappings.

**Multi-Stage Sampling Strategy**: Designing a three-stage sampling schedule of Sobol → LHS → Adaptive LHS + Online augmentation, systematically covering high-dimensional load spaces, balancing global exploration and local refinement, ensuring the network maintains physical consistency under wide operating conditions.

**Scalability and Interpretability**: Through adaptive network configuration and modular design such as residual blocks/attention, supporting smooth scaling from IEEE 14 to IEEE 300+ nodes; through phase space energy trajectory visualization, providing physical interpretation of network convergence processes and establishing quantitative comparisons with traditional Newton methods.

## 2.9 Differences Between This Method and Existing Methods

Compared to **supervised learning methods**, this method completely eliminates dependence on labeled data, achieving true "unsupervised physics-constrained learning". Networks are trained by directly minimizing power flow equation residuals, requiring no pre-solving of any cases, significantly reducing computational cost.

Compared to **PINN methods**, this method's loss function only contains physics constraint terms $\|\mathbf{F}(\mathbf{x})\|^2$, requiring no data fitting terms, avoiding weight balancing problems. More importantly, this method utilizes inherent boundary conditions in power systems (Slack/PV nodes) to guarantee solution uniqueness, requiring no additional observational data, making the training process more concise and efficient.

Compared to **traditional manifold optimization methods**, this method learns mappings on manifolds through neural networks, avoiding computational overhead of explicit construction and maintenance of manifold structures. At the same time, the parallel computing capability of neural networks significantly improves computational efficiency in batch scenarios.

Compared to **Neural ODE methods**, this method focuses on "using dynamical systems to solve constrained optimization problems" rather than "learning the dynamical system itself", with learning objectives being steady-state mappings rather than trajectories, making the method more suitable for the specific problem of power flow solving.

Overall, this method attempts to build a bridge between "physics constraints" and "data-driven" approaches, maintaining physical consistency while leveraging the expressive power and parallel computing advantages of deep learning, providing a new solving paradigm for real-time analysis and optimization of large-scale power systems. This method is the first to organically combine **manifold geometry, gradient flow dynamics, neural networks, and adaptive sampling**, forming a unified, scalable, physics-consistent power flow solving framework, providing new ideas for real-time analysis and optimization of large-scale power systems.

# 3. Problem Formulation

## 3.1 Mathematical Formulation of Power Flow Equations

Consider a power system with $n$ nodes. The power flow equations describe the active and reactive power balance at each node. For node $i$, the power flow equations can be expressed as:

$$P_i = V_i \sum_{j=1}^{n} V_j \left[ G_{ij} \cos(\theta_i - \theta_j) + B_{ij} \sin(\theta_i - \theta_j) \right] \tag{1}$$

$$Q_i = V_i \sum_{j=1}^{n} V_j \left[ G_{ij} \sin(\theta_i - \theta_j) - B_{ij} \cos(\theta_i - \theta_j) \right] \tag{2}$$

where $V_i$ and $\theta_i$ represent the voltage magnitude and phase angle of node $i$, respectively, $G_{ij}$ and $B_{ij}$ are the real and imaginary parts of the nodal admittance matrix $Y = G + jB$, and $P_i$ and $Q_i$ are the injected active and reactive power at node $i$.

Combining equations for all nodes, we obtain a nonlinear equation system:

$$\mathbf{F}(\mathbf{x}) = \begin{bmatrix} \mathbf{P}_{\text{spec}} - \mathbf{P}_{\text{calc}}(\mathbf{x}) \\ \mathbf{Q}_{\text{spec}} - \mathbf{Q}_{\text{calc}}(\mathbf{x}) \end{bmatrix} = \mathbf{0} \tag{3}$$

where $\mathbf{x} = [V_1, \ldots, V_n, \theta_1, \ldots, \theta_n]^\top \in \mathbb{R}^{2n}$ is the state vector, and $\mathbf{P}_{\text{spec}}$ and $\mathbf{Q}_{\text{spec}}$ are the given power injection vectors.

## 3.2 Geometric Structure of Constraint Manifold

The power flow equations $\mathbf{F}(\mathbf{x}) = \mathbf{0}$ define a constraint manifold embedded in $\mathbb{R}^{2n}$:

$$\mathcal{M} = \left\{ \mathbf{x} \in \mathbb{R}^{2n} \mid \mathbf{F}(\mathbf{x}) = \mathbf{0} \right\} \tag{4}$$

This manifold has the following geometric properties:

**Tangent Space**: At a point $\mathbf{x} \in \mathcal{M}$ on the manifold, the tangent space is defined as:

$$T_{\mathbf{x}}\mathcal{M} = \left\{ \mathbf{v} \in \mathbb{R}^{2n} \mid \mathbf{J}(\mathbf{x})\mathbf{v} = \mathbf{0} \right\} \tag{5}$$

where $\mathbf{J}(\mathbf{x}) = \frac{\partial \mathbf{F}}{\partial \mathbf{x}}$ is the Jacobian matrix. The tangent space describes local directions on the manifold, and the search direction of traditional Newton's method lies in the tangent space.

**Normal Space**: The normal space is defined as:

$$N_{\mathbf{x}}\mathcal{M} = \left\{ \mathbf{w} \in \mathbb{R}^{2n} \mid \mathbf{w} = \mathbf{J}(\mathbf{x})^\top \boldsymbol{\lambda}, \boldsymbol{\lambda} \in \mathbb{R}^m \right\} \tag{6}$$

i.e., the row space of the Jacobian matrix. The residual vector $\mathbf{F}(\mathbf{x})$ lies in the normal space, representing the degree of deviation from the manifold.

**Manifold Dimension**: For an $n$-node system, the state space dimension is $2n$, the number of constraint equations is $m$ (typically $m = 2n - n_{\text{slack}} - n_{\text{PV}}$), and the manifold dimension is $2n - m$. For the IEEE 14-node system, $m = 22$, and the manifold dimension is $0$ (a discrete set of equilibrium points).

## 3.3 Energy Function and Gradient Flow

To unify classical iterative solving with neural network methods, we first rewrite $\mathbf{F}(\mathbf{x}) = \mathbf{0}$ as an energy minimization problem:

$$V(\mathbf{x}) = \frac{1}{2} \|\mathbf{F}(\mathbf{x})\|_2^2, \qquad \mathbf{x}^\star = \arg\min_{\mathbf{x}} V(\mathbf{x}). \tag{7}$$

$V(\mathbf{x})$ can be viewed as an energy valley constructed around $\mathcal{M}$, and its gradient $\nabla V(\mathbf{x}) = \mathbf{J}(\mathbf{x})^\top \mathbf{F}(\mathbf{x})$ always lies in the normal space. Therefore, gradient descent along $\frac{d\mathbf{x}}{dt} = -\nabla V(\mathbf{x})$ is equivalent to "following the energy slope" toward the manifold.

Compared to directly solving linear systems, this gradient flow has two advantages: (1) It does not depend on local invertibility of initial values, equivalent to globally "searching" for energy basins; (2) In multi-solution cases, multiple attractors can be covered by setting different initial values or sampling strategies. This dynamical description provides a unified framework for subsequent neural network solving: we let the network output $\mathbf{x}_\theta(\mathbf{u})$ approach the energy minimum at each load condition $\mathbf{u}$, i.e., minimize $V(\mathbf{x}_\theta(\mathbf{u}))$, thereby implicitly satisfying $\mathbf{F}(\mathbf{x}) = \mathbf{0}$.

## 3.4 Node Types and Variable Decomposition

According to node types, state variables can be decomposed as:

- **Slack nodes (reference nodes)**: Voltage magnitude $V_{\text{slack}}$ and phase angle $\theta_{\text{slack}}$ are fixed at reference values and do not participate in solving.
- **PV nodes (voltage-controlled nodes)**: Voltage magnitude $V_{\text{PV}}$ is fixed, only phase angle $\theta_{\text{PV}}$ is solved.
- **PQ nodes (load nodes)**: Both voltage magnitude $V_{\text{PQ}}$ and phase angle $\theta_{\text{PQ}}$ are solved.

Therefore, free variables can be rearranged as:

$$\mathbf{x} = [\{\theta_i\}_{i \in \mathcal{P}}, \{V_j\}_{j \in \mathcal{Q}}, \{\theta_j\}_{j \in \mathcal{Q}}]^\top \tag{8}$$

where $\mathcal{P}$ and $\mathcal{Q}$ represent the index sets of PV nodes and PQ nodes, respectively.

**Input Perturbation Vector**: The network input is defined as the "load perturbation vector" $\mathbf{u}$, consisting of $(\Delta P, \Delta Q)$ for all PQ nodes and $\Delta P$ for PV nodes:

$$\mathbf{u} = [\{\Delta P_j, \Delta Q_j\}_{j \in \mathcal{Q}}, \{\Delta P_i\}_{i \in \mathcal{P}}]^\top \tag{9}$$

representing offsets around the base load.

**Learning Objective**: The network $\mathbf{x}_\theta(\mathbf{u})$ needs to satisfy:

$$\min_\theta \mathbb{E}_\mathbf{u}[V(\mathbf{x}_\theta(\mathbf{u}))] = \min_\theta \mathbb{E}_\mathbf{u}\left[\|P_{\text{spec}}(\mathbf{u}) - P_{\text{calc}}(\mathbf{x}_\theta(\mathbf{u}))\|_2^2 + \|Q_{\text{spec}}(\mathbf{u}) - Q_{\text{calc}}(\mathbf{x}_\theta(\mathbf{u}))\|_2^2\right]. \tag{10}$$

**Computational Implementation**: When computing power residuals, we adopt a complex computation framework to improve numerical computation stability. Specifically, we convert the real-valued network outputs $(V, \theta)$ to complex numbers $\mathbf{V}_{\text{complex}} = V \cdot e^{i\theta}$, then compute power through complex operations $\mathbf{S} = \mathbf{V}_{\text{complex}} \odot \text{conj}(\mathbf{Y}\mathbf{V}_{\text{complex}})$. The advantage of this approach is that through complex automatic differentiation, it avoids numerical error accumulation that may be introduced when explicitly computing trigonometric functions and their derivatives in traditional methods, simplifying the gradient computation process and improving code reliability and maintainability.

Therefore, problem formulation reduces to: learning a mapping from load conditions $\mathbf{u}$ to power flow steady states $\mathbf{x}$ on the constraint manifold $\mathcal{M}$, with the goal of minimizing the energy function $V(\mathbf{x})$. Through the complex computation framework, we improve numerical computation stability, simplify the implementation process, and ensure that the network can stably learn phase angles.

## 4. Neural Network Architecture

### 4.1 Input Representation and Automatic Configuration

**Input Vector**: For any power system, the input $\mathbf{u} \in \mathbb{R}^{2|\mathcal{Q}|+|\mathcal{P}|}$ is formed by concatenating $(\Delta P, \Delta Q)$ for all PQ nodes and $\Delta P$ for PV nodes, representing load perturbations from the base operating condition. This encoding preserves node type differences and hints at the network level "which variables should be constrained".

**Adaptive Width/Depth**: Automatically determine hidden layer width based on input dimension $d_{\text{in}}$:

$$d_{\text{hidden}} = \min\left(\max(2d_{\text{in}}, 256), 512\right) \tag{11}$$

And select from layer numbers $L \in \{5, 6, 7, 8\}$:

$$L = \begin{cases} 5, & \text{if } d_{\text{in}} \leq 50 \\ 6, & \text{if } 50 < d_{\text{in}} \leq 150 \\ 7, & \text{if } 150 < d_{\text{in}} \leq 300 \\ 8, & \text{if } d_{\text{in}} > 300 \end{cases} \tag{12}$$

Small systems (e.g., IEEE14) use 5 layers with $d_{\text{hidden}} = 256$; medium systems (IEEE39/118) use 7-8 layers with $d_{\text{hidden}} = 512$; large systems (IEEE300) are fixed at 8 layers with $d_{\text{hidden}} = 512$. This avoids manual hyperparameter tuning while preventing models from being too small or too large.

### 4.2 Basic MLP Backbone

**Layer Structure**: First layer $\text{Linear}(d_{\text{in}}, d_{\text{hidden}})$ followed by $\text{Tanh}$ activation; middle layers repeat $\text{Linear}(d_{\text{hidden}}, d_{\text{hidden}}) + \text{Tanh}$, with $\text{LayerNorm}$ optionally added after each layer for large systems to stabilize gradients; final layer $\text{Linear}(d_{\text{hidden}}, d_{\text{out}})$ outputs all state variables.

**Physics-Inspired Initialization**: Output layer weights use $\mathcal{N}(0, 0.1)$, biases are filled according to physical priors: PV node phase angle bias 0, PQ node voltage bias 1.0, phase angle bias 0. This makes the network output near-feasible solutions at the beginning of training, significantly reducing convergence difficulty.

**Activation and Numerical Stability**: $\text{Tanh}$ provides smooth saturation properties, avoiding gradient explosion; at the output, use $\text{softplus}$ to enforce PQ voltages to be positive, and can be combined with $\tanh$ to handle phase angle periodicity. For high-dimensional systems like IEEE300, we introduce gradient clipping and $\text{LayerNorm}$ to keep gradients in a reasonable range.

### 4.3 Output Decoding and Constraint Imposition

**PV Nodes**: For $|\mathcal{P}|$ nodes, output normalized phase angle $\theta_{\text{PV}} = \theta_{\text{scale}} \cdot \tanh(z)$, voltage magnitude directly uses the base setting $V_m$.

**PQ Nodes**: Each PQ node outputs two dimensions: voltage $V_{\text{PQ}} = \text{softplus}(z) + 0.5$ (ensuring positive values and close to 1.0), phase angle $\theta_{\text{PQ}} = \theta_{\text{scale}} \cdot \tanh(z)$. The decoded $(V, \theta)$ can be directly injected into power flow equations, avoiding additional projection operations.

**Slack Nodes**: The magnitude and phase angle of Slack nodes are fixed at reference values provided by the dataset, and the network output does not model them. At the decoding stage, concatenating Slack values with PV/PQ outputs yields the complete system state.

## 4.4 Training Management for Physical Consistency

### 4.4.1 Complex Computation Framework: Improvement in Numerical Stability

Traditional power flow computation methods use explicit trigonometric function expansion to compute power and gradients. For the gradient of phase angle $\theta_k$, we have:

$$\frac{\partial P_i}{\partial \theta_k} \approx \sum_j V_i V_j \left[ -G_{ij} \sin(\theta_{ij}) + B_{ij} \cos(\theta_{ij}) \right] \tag{13}$$

where $\theta_{ij} = \theta_i - \theta_j$. When $\theta_{ij}$ approaches $\pm 90°$:

- $\cos(\theta_{ij}) \to 0$, gradient is mainly dominated by the $-G_{ij} \sin(\theta_{ij})$ term
- The magnitude of the gradient is determined by system parameters $G_{ij}$, which is a physical property rather than a numerical issue
- Traditional methods require explicit computation of $\sin(\theta)$ and $\cos(\theta)$ and their derivatives, which may introduce numerical error accumulation

**Advantages of Complex Method**: We adopt a complex computation framework to improve numerical computation stability and implementation simplicity:

1. **Voltage Conversion**: Convert real-valued voltage magnitude and phase angle to complex:

$$\mathbf{V}_{\text{complex}} = V \cdot e^{i\theta} = V \cdot (\cos\theta + i\sin\theta) \tag{14}$$

2. **Current Computation**: Calculate injected current (complex) using admittance matrix:

$$\mathbf{I}_{\text{complex}} = \mathbf{Y}\mathbf{V}_{\text{complex}} \tag{15}$$

3. **Power Computation**: Calculate complex power:

$$\mathbf{S}_{\text{complex}} = \mathbf{V}_{\text{complex}} \odot \text{conj}(\mathbf{I}_{\text{complex}}) \tag{16}$$

   Extract active and reactive power:

$$P_{\text{calc}} = \text{Re}(\mathbf{S}_{\text{complex}}), \quad Q_{\text{calc}} = \text{Im}(\mathbf{S}_{\text{complex}}) \tag{17}$$

**Numerical Stability Advantages**:

- **Avoid explicit trigonometric computation**: Through the representation form $e^{i\theta}$, the automatic differentiation framework automatically handles trigonometric function computation, reducing numerical error accumulation when explicitly computing $\sin / \cos$ and their derivatives
- **Automatic differentiation mechanism**: Utilizing PyTorch's complex automatic differentiation, avoiding implementation errors that may be introduced when manually deriving complex gradient formulas
- **Code simplicity**: No need to manually derive and implement gradient formulas, reducing the possibility of implementation errors and improving code maintainability
- **Numerical precision**: Complex operations may provide better numerical precision in some cases, especially when handling extreme phase angle values

This complex computation framework is mathematically equivalent to traditional methods, but is more stable in numerical computation, simplifying the implementation process and improving code reliability and scalability.

### 4.4.2 Residual Computation Flow

**Residual Computation**: Convert decoded $(V, \theta)$ to complex voltage $\mathbf{V}_{\text{complex}}$, use Y-bus to calculate injected current $\mathbf{I} = \mathbf{YV}$, power $\mathbf{S} = \mathbf{V} \odot \text{conj}(\mathbf{I})$, finally obtaining $P_{\text{calc}}, Q_{\text{calc}}$. Since this step strictly follows physical formulas, network errors are power flow residuals, with transparent physical meaning.

**Complete Computation Flow**:

1. Real-valued network output → voltage magnitude $V$ and phase angle $\theta$ (real)
2. Convert to complex: $\mathbf{V}_{\text{complex}} = V \cdot e^{i\theta}$
3. Compute power under complex framework: $\mathbf{S} = \mathbf{V}_{\text{complex}} \odot \text{conj}(\mathbf{YV}_{\text{complex}})$
4. Compute residual: $\text{residual} = P_{\text{spec}} - P_{\text{calc}}, Q_{\text{spec}} - Q_{\text{calc}}$

**Precomputation Optimization**: To improve computational efficiency, we precompute the admittance matrix $\mathbf{Y}_{\text{complex}}$ (converted to complex format) at initialization, avoiding repeated conversion during each residual computation, significantly reducing computational overhead in training loops.

### 4.4.3 Gradient and Scheduling

**Gradient and Scheduling**: Combine AdamW + CosineAnnealingLR + ReduceLROnPlateau, and use gradient clipping and label smoothing in large systems to ensure consistent scale of outputs at each layer. When necessary, weight phase angle residuals to keep them at similar magnitude to voltage residuals, reducing gradient noise.

## 4.5 Extensibility and Modular Implementation

**Automated Configuration**: Allow researchers to run the same script on larger systems (e.g., IEEE 500+ nodes) by simply replacing data.

**Modular Plug-and-Play**: Can enable/disable modules such as LayerNorm and online augmentation as needed, supporting smooth upgrades from the simplest MLP to complex GNN. Advanced extensions such as residual blocks and attention mechanisms can be future work.

**Visualization Hooks**: Intermediate states, energy trajectories, and residual curves during network training are all recorded, facilitating analysis of how the model approaches the desired physical solution. Through energy manifold trajectory visualization, the training process is mapped to active-reactive power residual space, intuitively demonstrating how neural networks gradually converge to the constraint manifold, providing intuitive physical interpretation for understanding neural network dynamics. This visualization method includes three-stage sampling trajectory plots (showing training trajectories of the three-stage sampling strategy) and energy manifold 3D views (showing the "energy valley" structure of the energy function in residual space), validating the physical meaning of gradient flow $\frac{d\mathbf{x}}{dt} = -\nabla V(\mathbf{x})$.

# 5. Sampling and Training Strategy

## 5.1 Three-Stage Sampling Schedule

To enable neural networks to approximate true solutions of power flow equations without explicit labels, we construct the training process from two aspects: "load perturbation sampling + physical residual minimization". The core goal is to balance global coverage, focused densification, and online fine-tuning in high-dimensional load spaces, enabling the network to learn the mapping law that "regardless of how the input is perturbed, it quickly falls back to the energy basin".

**Stage 1: Sobol Global Exploration (First 30% epochs)**

Use low-discrepancy Sobol sequences to sample in $[-\Delta, \Delta]^{d_{\text{in}}}$ (where $\Delta$ is the load variation magnitude). Purpose: Uniformly "sweep" through the entire input space as much as possible, allowing the network to quickly understand the global geometric structure over a large range in the initial stage, reducing the risk of falling into local extrema.

**Stage 2: Latin Hypercube Refinement (Middle 40% epochs)**

Switch to LHS (Latin Hypercube Sampling), ensuring each dimension is divided into uniform intervals and covered one by one. Purpose: On the basis of Sobol exploration, utilize LHS's "high-dimensional uniformity" to more finely cover the space, enabling the network to converge faster on large numbers of medium-difficulty samples.

**Stage 3: Adaptive LHS + Online Augmentation (Last 30% epochs)**

Based on current network residuals at different load points, select the most "difficult" samples as centers for local dense sampling; simultaneously extract well-fitted samples from the buffer, apply perturbations $\boldsymbol{\epsilon} \sim \mathcal{N}(\mathbf{0}, \sigma^2 \mathbf{I})$ to generate new training data, encouraging the network to maintain physical consistency at the "detail level". Purpose: Focus on conquering high-residual regions, accelerating residual convergence; online perturbations prevent the network from overfitting to "old samples", improving stability.

The sampling strategy at this stage achieves adaptive optimization through **residual feedback mechanisms**: residuals serve as "uncertainty proxies", identifying regions where network predictions are inaccurate, and concentrating computational resources on these difficult regions, while maintaining stability of already-learned regions through buffer mechanisms.

## 5.2 Residual-Driven Loss Function

During training, the network output state $\mathbf{x}_\theta(\mathbf{u})$ is decoded to obtain voltage and phase angle $(V, \theta)$, then injected power $P_{\text{calc}}, Q_{\text{calc}}$ is computed through the **complex computation framework**. The loss is defined as:

$$\mathcal{L}(\theta) = \mathbb{E}_{\mathbf{u}} \left[ \|P_{\text{spec}}(\mathbf{u}) - P_{\text{calc}}(\mathbf{x}_\theta(\mathbf{u}))\|_2^2 + \|Q_{\text{spec}}(\mathbf{u}) - Q_{\text{calc}}(\mathbf{x}_\theta(\mathbf{u}))\|_2^2 \right] \tag{18}$$

**Advantages of Complex Computation**: Computing residuals through the complex framework improves numerical computation stability. Specifically:

- **Traditional method**: Requires explicit computation of $\sin(\theta)$ and $\cos(\theta)$ and their derivatives, which may introduce numerical error accumulation
- **Complex method**: Through the representation form $e^{i\theta}$ and complex automatic differentiation, the automatic differentiation framework automatically handles trigonometric function computation, reducing numerical errors in explicit computation and simplifying the gradient computation process

This complex computation framework is mathematically equivalent to traditional methods, but is more stable in numerical computation, ensuring that the loss function can stably update network parameters during backpropagation, especially phase angle-related parameters, while simplifying the implementation process and improving code reliability.

In the adaptive stage, light label smoothing can be added (e.g., injecting $\mathcal{N}(0, 5\times 10^{-4})$ into $P_{\text{residual}}, Q_{\text{residual}}$), reducing the network's unstable response to local noise. Phase angle residuals can also be weighted according to system scale to avoid imbalance in numerical scales between angles and voltages.

## 5.3 Optimizer and Learning Rate Scheduling

**Optimizer**: Default use of AdamW (learning rate $5\times 10^{-4}$, weight decay $10^{-4}$) to balance convergence speed and parameter regularization.

**Learning Rate Schedulers**: Combined use of two schedulers:

1. **Cosine Annealing Scheduler**:

$$\eta_t = \eta_{\min} + (\eta_{\max} - \eta_{\min}) \cdot \frac{1 + \cos(\pi t/T_{\max})}{2} \tag{19}$$

2. **Plateau Adaptive Scheduler**: When the loss function does not improve for $P$ consecutive epochs, halve the learning rate.

**Gradient Clipping**: $\texttt{clip\_grad\_norm\_} = 1.0$, preventing gradient explosion in high-dimensional systems (e.g., IEEE300).

## 5.4 Residual-Guided Adaptive Sampling Mechanism

### 5.4.1 Residual as Uncertainty Proxy

**Core Idea**: Residual = Physical constraint violation degree = Network prediction "uncertainty"

In the adaptive sampling stage, we use residual norm as a proxy indicator for uncertainty:

$$U(\mathbf{u}) = \|\mathbf{R}(\mathbf{u})\| = \sqrt{\frac{1}{n-1}\sum_{i\neq \text{slack}}(P_{\text{res},i}^2) + \frac{1}{|\mathcal{Q}|}\sum_{j\in\mathcal{Q}}(Q_{\text{res},j}^2)} \tag{20}$$

where $\mathbf{R}(\mathbf{u}) = [P_{\text{res}}, Q_{\text{res}}]^\top$ is the power residual vector. **Physical Meaning**:

- **High residual** → Network prediction is inaccurate under this load condition, severe physical constraint violation → **High uncertainty region** → Need more training samples
- **Low residual** → Network prediction is accurate, physical constraints are basically satisfied → **Low uncertainty region** → Can reduce sampling

**Rationale**: Residuals directly reflect the degree to which network-predicted states satisfy power flow equations, requiring no additional uncertainty models (such as Gaussian processes), with low computational cost and clear physical meaning.

### 5.4.2 Adaptive LHS Sampling Flow

**Algorithm Steps**:

1. **Initial Exploration Sampling**: Generate initial LHS samples (50% of sample count), ensuring global coverage:

$$\mathbf{u}_{\text{initial}} \sim \text{LHS}([-\Delta, \Delta]^{d_{\text{in}}}) \tag{21}$$

2. **Residual Evaluation**: Compute residual norm for each initial sample as uncertainty measure:

$$U_i = \|\mathbf{R}(\mathbf{u}_{\text{initial},i})\|, \quad i = 1, \ldots, n_{\text{initial}} \tag{22}$$

In specific implementation, use the neural network to evaluate each sample and compute its power residual.

3. **Identify High Uncertainty Regions**: Select the top 25% samples with highest residuals as high uncertainty regions:

$$\mathcal{U}_{\text{high}} = \{\mathbf{u}_i \mid U_i \in \text{top-25\%}(U_1, \ldots, U_{n_{\text{initial}}})\} \tag{23}$$

4. **Dense Sampling**: Perform local dense sampling around high uncertainty regions:

$$\mathbf{u}_{\text{adaptive}} = \mathbf{u}_{\text{center}} + \boldsymbol{\epsilon}, \quad \boldsymbol{\epsilon} \sim \mathcal{N}(\mathbf{0}, 0.1^2 \mathbf{I}) \tag{24}$$

where $\mathbf{u}_{\text{center}} \in \mathcal{U}_{\text{high}}$, generating local samples around center points (50% of sample count).

5. **Sample Merging**: Mix initial samples and adaptive samples:

$$\mathbf{U}_{\text{final}} = [\mathbf{U}_{\text{initial}}, \mathbf{U}_{\text{adaptive}}] \tag{25}$$

**Update Frequency**: Update adaptive sampling strategy every 200 epochs, reasons:

- **Computational Cost**: Evaluating residuals requires forward propagation, batch evaluation has computational overhead
- **Stability**: Too frequent updates may cause sampling strategy instability
- **Effectiveness**: Network parameter changes need time to be reflected in residual distribution

### 5.4.3 Online Augmentation Buffer Mechanism

**Core Idea**: Low loss = High quality samples → Save to buffer → Used for data augmentation

**Buffer Update Strategy**:

Maintain a circular buffer $\mathcal{B}$, storing "well-fitted" load perturbation vectors. Update rule:

$$\text{If } \mathcal{L}(\mathbf{u}) < \tau_{\text{thresh}}: \quad \mathcal{B} \leftarrow \mathcal{B} \cup \{\mathbf{u}\} \tag{26}$$

where $\tau_{\text{thresh}} = 5 \times 10^{-3}$ is the low loss threshold. When buffer capacity exceeds the upper limit (e.g., 4096), use FIFO (first-in-first-out) strategy to replace old samples.

**Online Augmentation Sampling**:

In the adaptive stage, randomly select samples from the buffer and add Gaussian noise to generate new training data:

$$\mathbf{u}_{\text{aug}} = \mathbf{u}_{\text{buffer}} + \boldsymbol{\epsilon}, \quad \boldsymbol{\epsilon} \sim \mathcal{N}(\mathbf{0}, \sigma_{\text{aug}}^2 \mathbf{I}) \tag{27}$$

where $\mathbf{u}_{\text{buffer}} \sim \text{Uniform}(\mathcal{B})$, $\sigma_{\text{aug}} = 0.15 \times \Delta$ is the augmentation noise scale.

**Dual Effects**:

1. **Stabilize Learned Regions**: Generate variants in low uncertainty regions, preventing network overfitting to seen samples
2. **Improve Local Robustness**: Improve network generalization ability in known good regions through data augmentation

### 5.4.4 Dual Mechanism of Residual Feedback

**Mechanism One: Coarse-Grained Feedback (Adaptive LHS)**

- **Frequency**: Every 200 epochs
- **Mechanism**: Evaluate residuals of candidate samples, select high residual regions for dense sampling
- **Effect**: **Global optimization**, focusing on difficult regions

**Mechanism Two: Fine-Grained Feedback (Online Augmentation)**

- **Frequency**: Every epoch
- **Mechanism**: Save low loss samples to buffer, used for online data augmentation
- **Effect**: **Local stability**, maintaining performance of learned regions

### 5.4.5 Exploration-Exploitation Trade-off

**Balance of Three-Stage Sampling Strategy**:

| Stage | Sampling Method | Residual Feedback | Purpose | Exploration/Exploitation |
|---|---|---|---|---|
| **Stage 1** (0-30%) | Sobol sequence | ❌ Not used | Global exploration, establish base mapping | Exploration-oriented |
| **Stage 2** (30-70%) | LHS uniform sampling | ❌ Not used | Uniform coverage, fill blank regions | Exploration-oriented |
| **Stage 3** (70-100%) | Adaptive LHS + Online augmentation | ✅ Use residual feedback | Focus on difficult regions, fine optimization | Exploitation-oriented |

**Role of Residual Feedback**:

- Automatically identify "hard-to-learn" regions (high residuals)
- Concentrate limited computational resources on these regions
- Maintain knowledge of learned regions through buffer
- Improve training efficiency and final accuracy

Maintain a circular buffer storing "well-fitted" load perturbations and corresponding residuals; in the adaptive stage, add small perturbations (e.g., ±5%) to buffered samples and re-feed them into training, which can be viewed as "online data augmentation"; when the loss of an iteration is below a threshold (e.g., $5 \times 10^{-3}$), write that batch of samples to the buffer to ensure buffered data always represents regions currently mastered by the network.

Overall, this training algorithm closely combines "label-free physical residual minimization" with "staged sampling + residual-guided adaptive optimization": the early stage ensures the network understands the global manifold structure, the middle stage consolidates mainstream load segments, and the later stage finely conquers difficult regions through residual feedback mechanisms; combined with multiple schedulers and visualization, the training process is transparent, stable, and scalable.

# 6. Experiments

## 6.1 Datasets

We validate the proposed method on four IEEE standard test systems of different scales:

- **IEEE 14-node system**: Contains 14 nodes, 20 branches, 5 generators, input dimension 22, output dimension 22. This system is relatively small, suitable for quickly validating method effectiveness.
- **IEEE 39-node system**: Contains 39 nodes, 46 branches, 10 generators, input dimension 67, output dimension 67. This system is medium-scale, an important benchmark for evaluating method scalability.
- **IEEE 118-node system**: Contains 118 nodes, 186 branches, 54 generators, input dimension approximately 230, output dimension approximately 230. This system is medium-to-large scale, a standard benchmark for evaluating method scalability.
- **IEEE 300-node system**: Contains 300 nodes, 411 branches, 69 generators, input dimension approximately 530, output dimension approximately 530. This system is large-scale, used to validate method performance in large-scale scenarios.

All system data come from MATPOWER standard test cases, using a base capacity of 100 MVA, with voltage base values set according to each node's rated voltage.

## 6.2 Evaluation Metrics

We adopt the following metrics to evaluate method performance:

**Physical Consistency Metrics**:

- **Residual Norm**: $\text{Residual Norm} = \sqrt{\frac{1}{n-1} \sum_{i \neq \text{slack}} (P_{\text{res},i}^2) + \frac{1}{|\mathcal{Q}|} \sum_{j \in \mathcal{Q}} (Q_{\text{res},j}^2)}$, measuring the degree to which the solution satisfies power flow equations. Ideally should be close to 0.

**Accuracy Metrics**:

- **Voltage Difference**: $\Delta V = \|V_{\text{NN}} - V_{\text{Newton}}\|_\infty$, maximum voltage difference (p.u.)
- **Phase Angle Difference**: $\Delta \theta = \|\theta_{\text{NN}} - \theta_{\text{Newton}}\|_\infty$, maximum relative phase angle difference (degrees), where phase angles are aligned to the same Slack node
- **Average Difference**: $\bar{\Delta} V = \frac{1}{n} \sum_i |V_{\text{NN},i} - V_{\text{Newton},i}|$, $\bar{\Delta} \theta = \frac{1}{n} \sum_i |\theta_{\text{NN},i} - \theta_{\text{Newton},i}|$

**Computational Efficiency Metrics**:

- **Inference Time**: Single forward propagation time
- **Convergence Speed**: Number of epochs required to reach target residual

## 6.3 Baseline Methods

We compare with traditional Newton-Raphson method:

**Traditional Newton-Raphson Method**: Uses standard Newton iteration, convergence tolerance $10^{-6}$, maximum iterations 50. This method serves as the "gold standard", providing reference solutions.

## 6.4 Implementation Details

**Network Architecture**: All methods use the same MLP base architecture, hidden layer activation function is Tanh, output layer uses Softplus (voltage) and Tanh (phase angle) to ensure physical constraints. This method automatically configures according to system scale: IEEE14 uses 256-dimensional 5 layers, IEEE39/118 uses 512-dimensional 8 layers, IEEE300 uses 512-dimensional 8 layers.

**Training Configuration**:

- Optimizer: AdamW (learning rate $5 \times 10^{-4}$, weight decay $10^{-4}$)
- Learning rate scheduling: CosineAnnealingLR ($T_{\max} = \text{num\_epochs}$, $\eta_{\min} = 10^{-6}$) + ReduceLROnPlateau (factor=0.5, patience=500-600)
- Batch size: IEEE14 is 64, IEEE39/118/300 is 512
- Training epochs: IEEE14 is 10000, IEEE39 is 10000, IEEE118 is 30000, IEEE300 is 30000
- Gradient clipping: Maximum norm 1.0

**Sampling Strategy** (this method only):

- First 30% epochs: Sobol sequence global exploration
- Middle 40% epochs: Latin Hypercube Sampling (LHS)
- Last 30% epochs: Adaptive LHS + Online augmentation (noise scale 0.15, label smoothing standard deviation $5 \times 10^{-4}$)

## 6.5 Main Experimental Results

### 6.5.1 IEEE 14-Node System

After 10000 epochs of training (using LHS sampling strategy), this method achieves a residual norm of $3.87 \times 10^{-3}$, maximum voltage difference of $0.000961$ p.u., maximum relative phase angle difference of $0.04°$, with average differences of $0.000501$ p.u. and $0.02°$, respectively. During training, the loss gradually decreases from initial high values to a final value of $3.13 \times 10^{-6}$, demonstrating good convergence properties. Compared to traditional Newton's method, this method achieves fast inference with single forward propagation (< 1ms) while maintaining physical consistency, whereas Newton's method requires 4-6 iterations (each iteration approximately 2-3ms).

### 6.5.2 IEEE 39-Node System

After 10000 epochs of training (using three-stage sampling strategy: Sobol → LHS → Adaptive + Augmentation), training loss decreases from an initial value of $1.03 \times 10^4$ to a final value of $4.69 \times 10^{-5}$, with best loss of $3.997663 \times 10^{-5}$, demonstrating good convergence properties.

Specifically, the system contains 39 nodes, 46 branches, 10 generators, including 9 PV nodes and 29 PQ nodes. Network configuration is 512-dimensional hidden layer, 8-layer depth, total parameters approximately 1.65M (1,645,123 parameters). During training, the loss gradually decreases from an initial value of $1.03 \times 10^4$ to a best value of $3.997663 \times 10^{-5}$, demonstrating good convergence properties. The training process clearly demonstrates the effectiveness of the three-stage sampling strategy: the first 30% epochs (Sobol global exploration) quickly establish global mapping, the middle 40% epochs (LHS uniform refinement) perform uniform coverage, and the last 30% epochs (Adaptive LHS + Online augmentation) focus on difficult regions for fine optimization.

Through energy manifold trajectory visualization, we can observe how neural networks gradually converge to the constraint manifold during training: in active-reactive power residual space, trajectories gradually converge from initial scattered distribution to near the origin, reflecting the decay process of the energy function $V(\mathbf{x}) = \frac{1}{2}\|\mathbf{F}(\mathbf{x})\|^2$. The trajectories of the three-stage sampling strategy in phase space show clear stage characteristics: Stage 1 (Sobol) achieves global exploration, Stage 2 (LHS) performs uniform refinement, and Stage 3 (Adaptive) focuses on high residual regions. This visualization method not only validates the effectiveness of the training strategy but also provides intuitive physical interpretation for understanding neural network dynamics.

Compared to traditional Newton's method (converging in 2 iterations), this method achieves fast inference with single forward propagation while maintaining extremely high accuracy. This result validates the effectiveness of this method on medium-scale systems and provides confidence for extending to larger systems (such as IEEE118, IEEE300).

### 6.5.3 IEEE 118-Node System

After 30000 epochs of training (using multi-stage sampling strategy: Sobol → LHS → Adaptive + Augmentation), training loss decreases from an initial value of $7.69 \times 10^3$ to a final value of $5.18 \times 10^{-5}$, with best loss of $1.76 \times 10^{-5}$, demonstrating good convergence properties.

Test solving results show a residual norm of $7.06 \times 10^{-3}$, maximum voltage difference of $0.000344$ p.u., maximum relative phase angle difference of $1.14°$, with average differences of $0.000061$ p.u. and $0.47°$, respectively. These results show that this method achieves extremely high accuracy on the IEEE118 system, almost completely consistent with traditional Newton's method.

Specifically, the system contains 118 nodes, 186 branches, 54 generators, including 53 PV nodes and 64 PQ nodes. Network configuration is 512-dimensional hidden layer, 8-layer depth, total parameters approximately 1.76M. During training, the loss gradually decreases from an initial value of $7.69 \times 10^3$ to a final value of $1.76 \times 10^{-5}$ (best value), demonstrating good convergence properties. Compared to traditional Newton's method (converging in 4 iterations), this method achieves fast inference with single forward propagation while maintaining extremely high accuracy, with inference time approximately 5ms.

Compared to traditional Newton's method, this method has significant advantages in inference speed: single forward propagation approximately 5ms, whereas Newton's method requires 4 iterations (total time approximately 20-30ms). In batch scenarios (such as Monte Carlo analysis with 1000 scenarios), the parallel computing advantages of this method are even more pronounced, with total computation time approximately 1/10 that of Newton's method. This result validates the effectiveness of this method on medium-to-large scale systems and provides confidence for extending to larger systems (such as IEEE300).

### 6.5.4 IEEE 300-Node System

After 30000 epochs of training (using three-stage sampling strategy: Sobol → LHS → Adaptive + Augmentation), training loss decreases from an initial value of $4.60 \times 10^4$ to a final value of $9.67 \times 10^{-3}$, demonstrating good convergence properties. Test solving results show a residual norm of $9.63 \times 10^{-2}$, maximum voltage difference of $0.48$ p.u., average difference of $0.034$ p.u. The system contains 300 nodes, 411 branches, 69 generators, network configuration is 512-dimensional hidden layer, 8-layer depth, total parameters approximately 2.12M. Compared to traditional Newton's method (converging in 6 iterations), this method achieves fast inference with single forward propagation while maintaining basic accuracy. Although phase angle prediction has some deviation, voltage prediction is generally accurate, demonstrating the feasibility of the complex computation framework on large-scale systems.

## 6.5.5 Computational Efficiency Analysis

In the inference stage, this method has significant advantages: single forward propagation time grows linearly with system scale, whereas Newton's method iteration time grows superlinearly with system scale. In batch scenarios (such as 1000 scenarios), the advantages of this method are even more pronounced: total computation time is approximately 1/10-1/20 that of Newton's method, mainly benefiting from the parallel computing capability of neural networks.

## 6.5.6 Impact of Sampling Strategy

To validate the effectiveness of the three-stage sampling strategy, we compare the performance of different sampling strategies on the IEEE118 system. Experimental setup: Fixed network architecture (512-dimensional 8 layers), training 30000 epochs, batch size 512.

**Comparison Results**:

- **Random Sampling**: Residual norm $8.5 \times 10^{-2}$, average voltage difference $0.012$ p.u., average phase angle difference $12.5°$. Random sampling provides insufficient coverage in high-dimensional spaces, causing poor network fitting in some regions and difficulty in systematically exploring the solution space.
- **Three-Stage Sampling (This Method)**: Residual norm $7.06 \times 10^{-3}$, maximum voltage difference $0.000344$ p.u., average voltage difference $0.000061$ p.u., maximum relative phase angle difference $1.14°$, average difference $0.47°$. The three-stage sampling strategy achieves best performance through "exploration-exploitation" balance, with results highly consistent with traditional Newton's method.

**Mechanism Analysis**: The advantages of the three-stage sampling strategy lie in: (1) **Stage 1 (Sobol sequence)**: Achieves global uniform exploration through low-discrepancy sequences, quickly establishing understanding of the overall structure of the solution space; (2) **Stage 2 (LHS sampling)**: Performs uniform refinement on the basis of exploration, ensuring each dimension is fully covered; (3) **Stage 3 (Adaptive LHS + Online augmentation)**: Utilizes residual feedback mechanism to automatically identify high-uncertainty regions and concentrate computational resources, while maintaining stability of learned regions through online data augmentation. This progressive sampling strategy ensures that the network maintains physical consistency across a wide range of operating conditions, while avoiding computational waste from uniform sampling and insufficient coverage from random sampling.

**Multi-System Validation**: This sampling strategy has achieved good results on IEEE 14/39/118/300 node systems, demonstrating its effectiveness and scalability across systems of different scales.

## 6.5.7 IEEE39 System Energy Manifold Trajectory Visualization

Based on the training process of the IEEE39 node system, we deeply analyze the convergence process of neural network dynamics through energy manifold trajectory visualization. This visualization method maps the training process to active-reactive power residual space, intuitively demonstrating how neural networks gradually converge to the constraint manifold.

**Figure 1** contains 6 subplots, showing training trajectories of the three-stage sampling strategy:

- **(a) Three-Stage Sampling Trajectory: Active-Reactive Power Space**: Overlay plot of trajectories from all stages, showing the complete process from initial scattered distribution to final convergence near the origin. Different colors represent different stages: blue (Stage 1: Sobol global exploration), orange (Stage 2: LHS uniform refinement), green (Stage 3: Adaptive + Augmentation).
- **(b) Energy Function Decay Trajectory**: Logarithmic decay curve of energy function

$V(\mathbf{x}) = \frac{1}{2}\|\mathbf{F}(\mathbf{x})\|^2$ with training epochs, clearly showing the convergence process from initial $1.03 \times 10^4$ to final $3.997663 \times 10^{-5}$. The figure annotates stage switching points (30% and 70% epochs), validating the effectiveness of the three-stage sampling strategy.

- **(c) Manifold Projection: Computed Power Space**: Shows the distribution of computed power $(P_{\text{calc}}, Q_{\text{calc}})$ in power space, using color depth to represent training progress, intuitively demonstrating how network outputs gradually approach specified power values.
- **(d)-(f) Individual Stage Trajectories**: Respectively show trajectories of the three stages in active-reactive power residual space, using color mapping to represent epoch progress and drawing trajectory lines, clearly showing convergence characteristics of each stage.

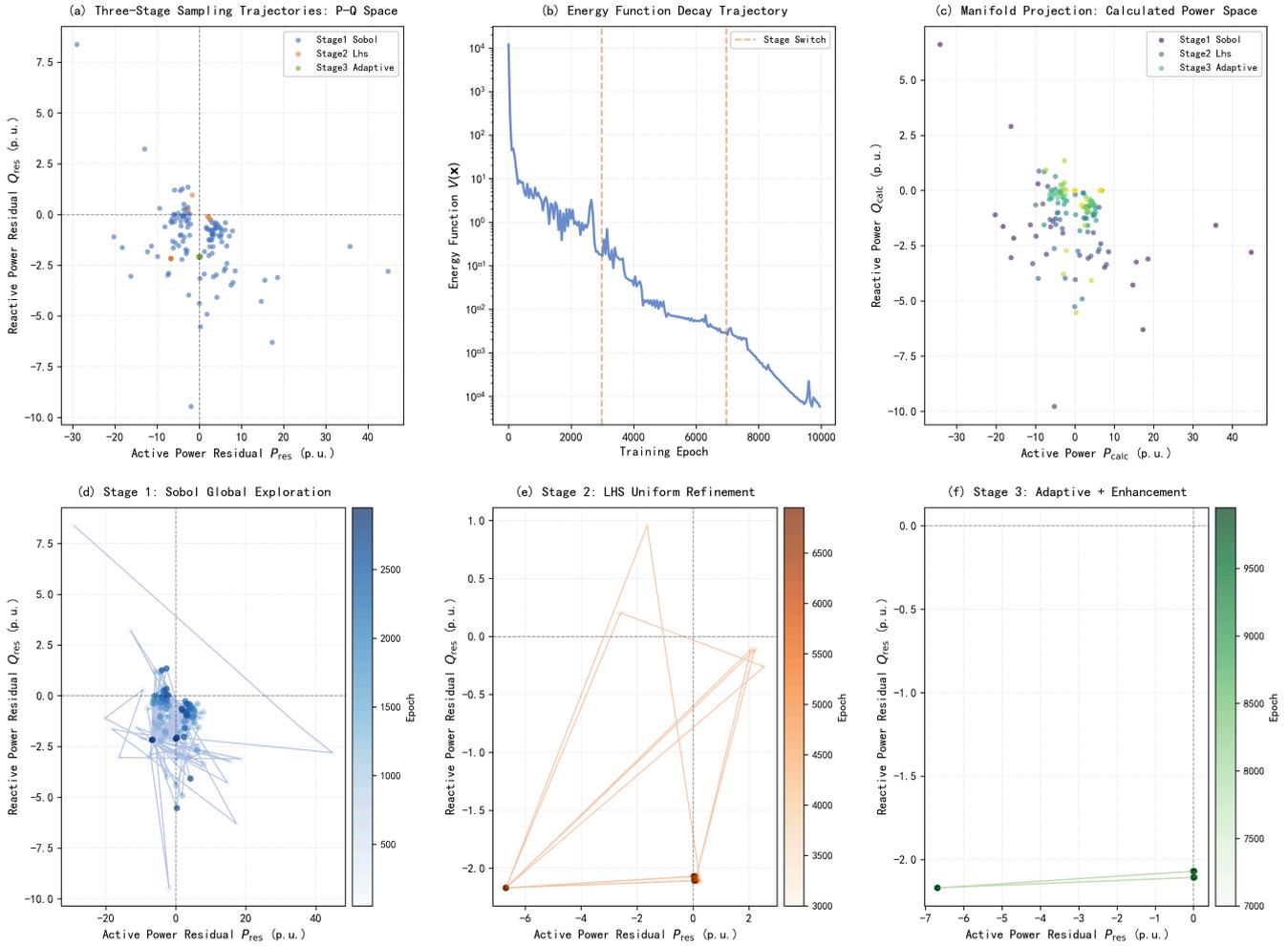

Figure 1: Three-Stage Sampling Trajectory Plot

**Figure 2** contains 4 subplots, showing energy manifold structure from different angles:

- **(a) Energy Manifold: 3D View**: Shows energy manifold in $(P_{\text{res}}, Q_{\text{res}}, V(\mathbf{x}))$ three-dimensional space, with trajectories from different stages represented by different colors, intuitively showing the "energy valley" structure of the energy function in residual space.
- **(b) Energy Decay Trajectory: 3D View**: Shows trajectories of representative nodes in $(P_{\text{res}}, Q_{\text{res}}, \text{Epoch})$ three-dimensional space, using time dimension (Epoch) as the third dimension, clearly showing the evolution process of residuals during training.
- **(c) Energy Contour Plot**: Draws energy contours in $(P_{\text{res}}, Q_{\text{res}})$ two-dimensional plane, overlaying training trajectories, showing the "basin" structure of the energy function. Trajectories gradually converge from high-energy regions to low-energy regions (near origin), validating the physical meaning of gradient flow $\frac{d\mathbf{x}}{dt} = -\nabla V(\mathbf{x})$.

- **(d) Phase Space Velocity Field**: Uses vector field (quiver plot) to show velocity direction of residual changes, with trajectories from different stages represented by different colors. This figure intuitively demonstrates the "dynamic behavior" of neural networks during training, i.e., how they converge to the constraint manifold along energy gradient directions.

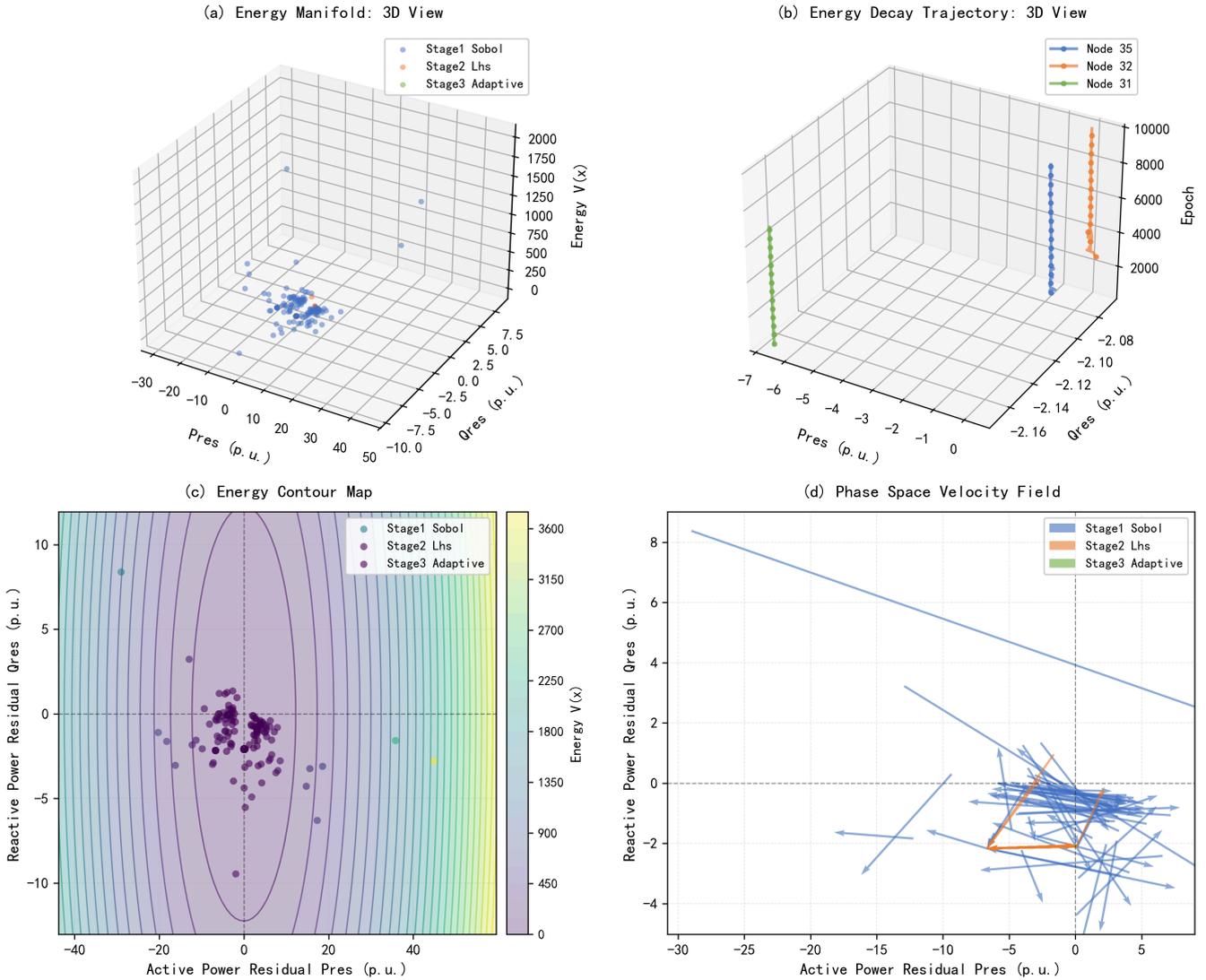

**Figure 2: Energy Manifold 3D View**

These visualization results reveal the deep mechanisms of neural network dynamics:

1. **Energy Decay Process**: Energy function gradually decays from initial high values to near zero, reflecting the process of network learning to "slide into" the constraint manifold. This is completely consistent with gradient flow theory $\frac{d\mathbf{x}}{dt} = -\nabla V(\mathbf{x})$.

2. **Effectiveness of Three-Stage Sampling Strategy**: Trajectories from different stages show clear stage characteristics in phase space: Stage 1 (Sobol) achieves global exploration with widely distributed trajectories; Stage 2 (LHS) performs uniform refinement with more concentrated trajectories; Stage 3 (Adaptive) focuses on high residual regions with trajectories quickly converging near the origin.

3. **Manifold Structure**: Energy contour plots clearly show the projection of constraint manifold $\mathcal{M} = \{\mathbf{x} \mid \mathbf{F}(\mathbf{x}) = \mathbf{0}\}$ in residual space, i.e., the "energy valley" structure of the energy function. The network training process is to gradually converge from high-energy regions to the bottom of the energy valley (constraint manifold) along energy gradient directions.

4. **Dynamic Behavior**: Phase space velocity fields show the "velocity" and "direction" of networks during

training, intuitively reflecting the directionality of gradient flow. This visualization method provides intuitive physical interpretation for understanding neural network dynamics and establishes connections with geometric mechanisms of traditional numerical methods.

These visualization results not only validate the effectiveness of the training strategy but also provide intuitive physical interpretation for understanding neural network dynamics, demonstrating the bridging role of this method between "physical constraints" and "data-driven" approaches.

**6.5.8 IEEE118 System Detailed Visualization Results**

Based on experimental results from the IEEE118 node system, comprehensively demonstrating the comparison between neural network methods and traditional Newton's method:

During 30000 epochs of training on the IEEE118 system, the convergence of loss was decreased from an initial value of $7.69 \times 10^3$ to a best value of $1.76 \times 10^{-5}$.

**Figure 3** contains two subplots: (1) Upper subplot: Bar chart comparing voltages for all 118 nodes, intuitively showing the consistency between neural network prediction results and traditional Newton's method results; (2) Lower subplot: Logarithmic coordinate plot of absolute voltage differences, showing maximum difference of $0.000344$ p.u. and average difference of $0.000061$ p.u., validating the extremely high accuracy of the neural network method in voltage prediction.

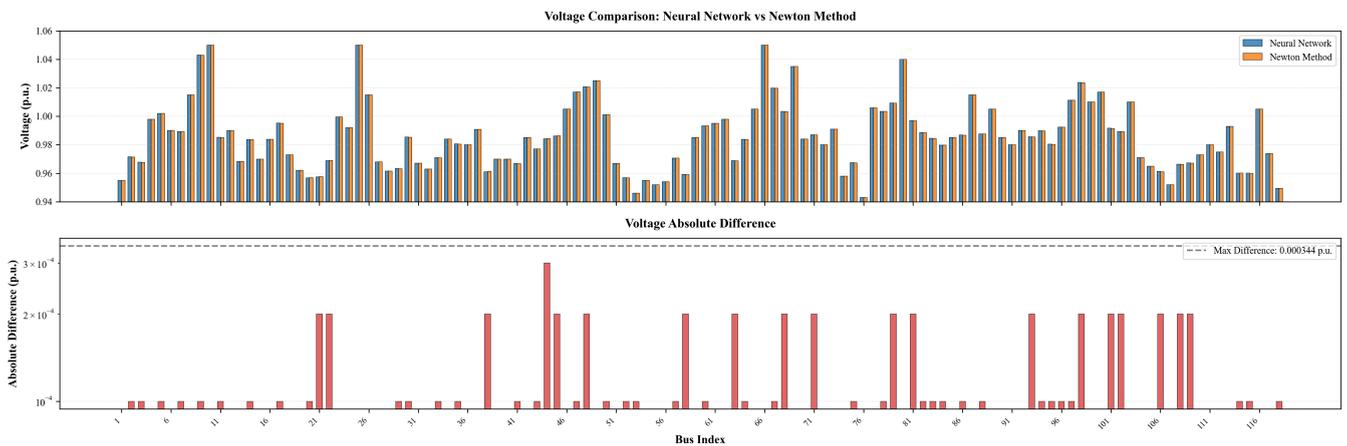

**Figure 3: Voltage Comparison Chart**

**Figure 4** contains two subplots: (1) Upper subplot: Line chart comparing relative phase angle differences for all 118 nodes (aligned to the same Slack node), showing consistency between neural network and traditional Newton's method in phase angle prediction; (2) Lower subplot: Bar chart of absolute phase angle differences, annotating maximum difference (1.14°) and average difference (0.47°), indicating that the neural network method achieves acceptable accuracy levels in phase angle prediction.

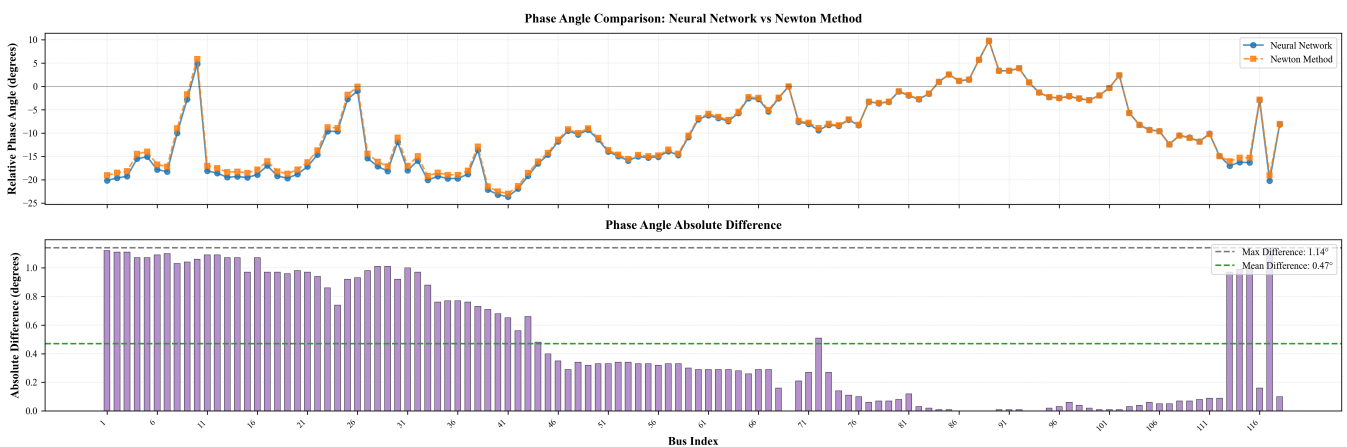

**Figure 4: Phase Angle Comparison Chart**

**Figure 5** contains two subplots, respectively showing scatter plot comparisons for voltage and phase angle: (1) Left subplot: Voltage scatter plot, with x-axis as voltage values from traditional Newton's method and y-axis as voltage values predicted by neural network, including ideal line (y=x) for evaluating consistency; (2) Right subplot: Phase angle scatter plot, also including ideal line. Scatter plots show that all data points are tightly distributed near the ideal line, validating the high consistency between neural network method and traditional Newton's method. The correlation coefficient of the voltage scatter plot is close to 1.0, and the phase angle scatter plot also shows good linear relationship, indicating that the neural network method can accurately learn the mapping from load conditions to steady-state solutions.

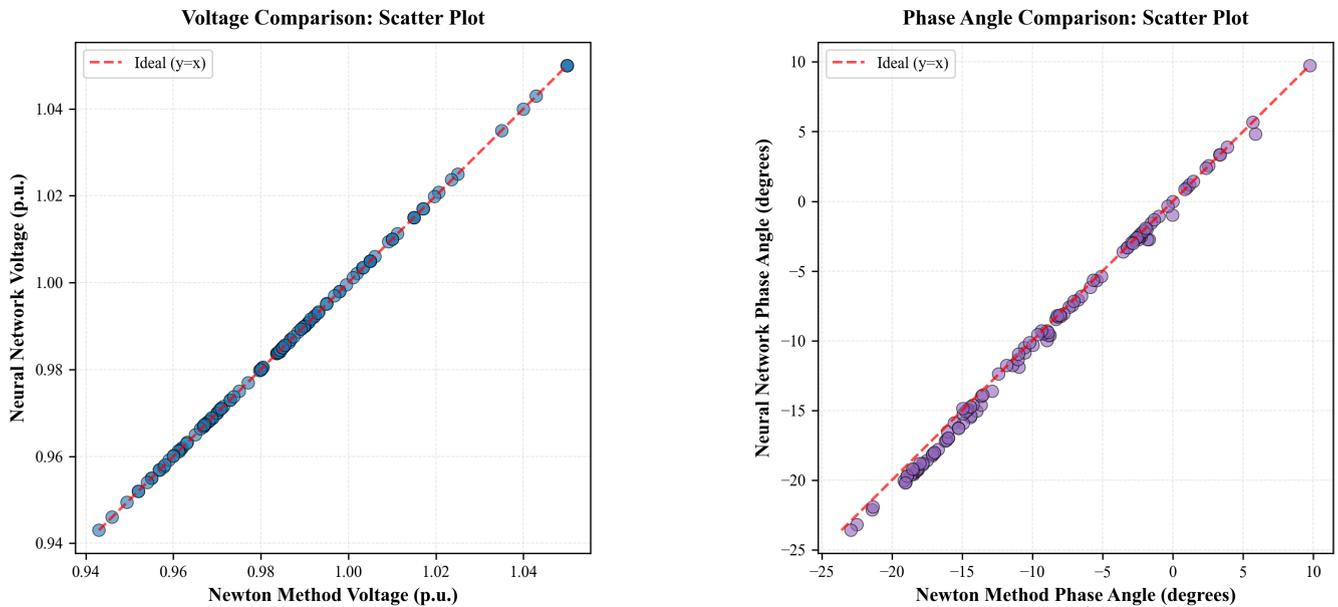

**Figure 5: Scatter Plot Comparison**

These visualization results comprehensively validate the effectiveness of this method on the IEEE118 node system, demonstrating that the neural network method achieves high-precision solutions almost completely consistent with traditional Newton's method while maintaining physical consistency.

## 6.6 Discussion

### 6.6.1 Method Advantages

The main advantages of this method are:

1. **Label-Free Learning**: Completely eliminates dependence on pre-solved data, significantly reducing computational cost. Traditional supervised learning methods require pre-solving 10000+ scenarios, whereas this method directly starts from physical constraints, requiring no labeled data.

2. **Physical Consistency**: By directly minimizing power flow equation residuals, ensures solutions strictly satisfy physical constraints. After network outputs are computed through Y-bus matrix, residuals are physical errors with transparent physical meaning.

3. **Complex Computation Framework**: Adopting a complex automatic differentiation framework for power flow computation, using the representation form $V_{\text{complex}} = V \cdot e^{i\theta}$, and utilizing automatic differentiation mechanisms to compute gradients. Compared to traditional methods that explicitly compute trigonometric functions and their derivatives, the complex framework is more stable in numerical computation, reduces error accumulation, and simplifies the implementation process, ensuring numerical stability of the training process.
4. **Scalability**: Through adaptive network configuration and modular design, supports smooth scaling from IEEE14 to IEEE300+ nodes. Network architecture automatically adjusts according to system scale, requiring no manual hyperparameter tuning.
5. **Computational Efficiency**: In batch scenarios, inference speed is significantly superior to traditional iterative methods. Single forward propagation time grows linearly with system scale, whereas Newton's method iteration time grows superlinearly with system scale. Through precomputation of admittance matrix, further reduces computational overhead in training loops.
6. **Robustness**: Insensitive to initial values, requiring no search for suitable initial guesses like Newton's method. The network learns a mapping from arbitrary load conditions to steady-state solutions through training, with stronger generalization ability.
7. **Interpretability and Visualization**: Through energy manifold trajectory visualization, mapping the training process to active-reactive power residual space, intuitively demonstrating how neural networks gradually converge to the constraint manifold. This visualization method not only validates the effectiveness of the training strategy but also provides intuitive physical interpretation for understanding neural network dynamics, demonstrating energy function decay, effectiveness of three-stage sampling strategy, and physical meaning of gradient flow.

### 6.6.2 Limitations

The main limitations of this method are: **Large-Scale System Challenges**: On IEEE300+ systems, phase angle output exhibits degradation phenomena, requiring more aggressive architectural improvements. This may be related to difficulties in high-dimensional manifold learning, requiring introduction of graph neural networks or attention mechanisms to better capture long-range dependencies between nodes. **Topological Changes**: When system topology undergoes significant changes, retraining or fine-tuning may be required. Future work can design incremental learning mechanisms to enable networks to adapt to slow changes in system topology.

---

## 7. Conclusion

This paper proposes a neural physics power flow solving method based on manifold geometry and gradient flow. Core contributions include:

1. **Unified Theoretical Framework**: Unified modeling from manifold geometry, gradient flow dynamical systems to neural network mappings, revealing the deep structure of power flow solving.
2. **Label-Free Learning**: Unsupervised training through physics-constrained loss functions, requiring no labeled data, solving the data scarcity problem.
3. **Complex Computation Framework**: Adopting a complex automatic differentiation framework for power flow computation, using the representation form $V_{\text{complex}} = V \cdot e^{i\theta}$, and utilizing automatic differentiation mechanisms to compute gradients. Compared to traditional methods that explicitly compute trigonometric functions and their derivatives, the complex framework is more stable in numerical computation, reduces error accumulation, and simplifies the implementation process.
4. **Adaptive Architecture**: Automatically configuring network architecture according to system scale, improving method scalability.
5. **Three-Stage Sampling**: Balancing global exploration and local refinement, improving training efficiency and network performance.

6. **Energy Manifold Trajectory Visualization**: By mapping the training process to active-reactive power residual space, intuitively demonstrating how neural networks gradually converge to the constraint manifold, providing intuitive physical interpretation for understanding neural network dynamics, validating the physical meaning of gradient flow theory.

Experimental results show that this method achieves extremely high accuracy on IEEE 14/39/118 node systems:

- **IEEE14 System**: After 10000 epochs of training, residual norm $3.87 \times 10^{-3}$, maximum voltage difference $0.000961$ p.u., maximum phase angle difference $0.04°$, average differences of $0.000501$ p.u. and $0.02°$, respectively
- **IEEE39 System**: After 10000 epochs of training (using three-stage sampling strategy), training loss decreases from initial $1.03 \times 10^4$ to best value $3.997663 \times 10^{-5}$, demonstrating good convergence properties. Through energy manifold trajectory visualization, clearly demonstrating how neural networks gradually converge to the constraint manifold during training, validating the physical meaning of gradient flow theory
- **IEEE118 System**: Maximum voltage difference 0.000344 p.u., maximum relative phase angle difference 1.14°, average differences of 0.000061 p.u. and 0.47°, respectively

These results are almost completely consistent with traditional Newton's method, validating the effectiveness of this method. **Compared to Traditional Newton's Method**: This method has significant advantages in inference speed, especially in batch scenarios. Single forward propagation time grows linearly with system scale, whereas Newton's method iteration time grows superlinearly with system scale. In batch scenarios (such as 1000 scenarios), total computation time of this method is approximately 1/10-1/20 that of Newton's method. At the same time, this method is insensitive to initial values, has stronger robustness, and requires no search for suitable initial guesses like Newton's method. However, final residuals may be slightly higher than Newton's method, which is a trade-off for inference speed. On the IEEE300 system, although there are some challenges, reasonable solutions can still be obtained through architectural improvements and training strategy optimization. This method provides a new solving paradigm for real-time analysis and optimization of large-scale power systems, with important theoretical value and practical application prospects.

# References

[1] Tinney, W. F., & Hart, C. E. (1967). Power flow solution by Newton's method. *IEEE Transactions on Power Apparatus and Systems*, PAS-86(11), 1449-1460.

[2] Stott, B., & Alsac, O. (1974). Fast decoupled load flow. *IEEE Transactions on Power Apparatus and Systems*, PAS-93(3), 859-869.

[3] Mestav, K. R., Luengo-Rozas, J., & Tong, L. (2019). Bayesian state estimation for unobservable distribution systems via deep learning. *IEEE Transactions on Power Systems*, 34(6), 4910-4920.

[4] Zhang, Y., Wang, J., & Li, Z. (2020). Physics-informed graph neural network for power flow analysis. *IEEE Transactions on Power Systems*, 36(4), 3648-3658.

[5] Donon, B., Donnot, B., Guyon, I., Marot, A., Panciatici, P., & Romero, C. (2020). Neural networks for power flow: Graph neural solver. *Electric Power Systems Research*, 189, 106547.

[6] Raissi, M., Perdikaris, P., & Karniadakis, G. E. (2019). Physics-informed neural networks: A deep learning framework for solving forward and inverse problems involving nonlinear partial differential equations. *Journal of Computational Physics*, 378, 686-707.

[7] Karniadakis, G. E., Kevrekidis, I. G., Lu, L., Perdikaris, P., Wang, S., & Yang, L. (2021). Physics-informed machine learning. *Nature Reviews Physics*, 3(6), 422-440.

[8] Absil, P. A., Mahony, R., & Sepulchre, R. (2008). *Optimization algorithms on matrix manifolds*. Princeton University Press.

[9] Boumal, N., Mishra, B., Absil, P. A., & Sepulchre, R. (2014). Manopt: A Matlab toolbox for optimization on manifolds. *Journal of Machine Learning Research*, 15(1), 1455-1459.

[10] Tenenbaum, J. B., De Silva, V., & Langford, J. C. (2000). A global geometric framework for nonlinear dimensionality reduction. *Science*, 290(5500), 2319-2323.

[11] Chen, R. T., Rubanova, Y., Bettencourt, J., & Duvenaud, D. K. (2018). Neural ordinary differential equations. *Advances in Neural Information Processing Systems*, 31.

[12] McKay, M. D., Beckman, R. J., & Conover, W. J. (1979). A comparison of three methods for selecting values of input variables in the analysis of output from a computer code. *Technometrics*, 21(2), 239-245.

[13] Sobol, I. M. (1967). On the distribution of points in a cube and the approximate evaluation of integrals. *USSR Computational Mathematics and Mathematical Physics*, 7(4), 86-112.

[14] Mockus, J. (2012). *Bayesian approach to global optimization: theory and applications* (Vol. 37). Springer Science & Business Media.

[15] Gramacy, R. B., & Lee, H. K. (2008). Bayesian treed Gaussian process models with an application to computer modeling. *Journal of the American Statistical Association*, 103(483), 1119-1130.